\documentclass[10pt,preprint2]{aastex}
\usepackage{latexsym}
\usepackage{graphics}

\begin{document}

\def\HI{\mbox{\ion{H}{1}}}
\def\HII{\mbox{\ion{H}{2}}}
\def\dim#1{\mbox{\,#1}}
\def\lya{\mbox{Ly-$\alpha$}}
\def\hide#1{}

\title{Redshifted 21 cm Emission From the Pre-Reionization Era\\
 II. \HII\ Regions Around Individual Quasars}
\author{Katharina Kohler, Nickolay Y.\ Gnedin}
\affil{CASA, University of Colorado, Boulder, CO 80309, USA;
kohlerk,gnedin@casa.colorado.edu}
\author{Jordi Miralda-Escud\'e}
\affil{Department of Astronomy, Ohio State University, Columbus, OH 43210;
 jordi@astronomy.ohio-state.edu} 
\and
\author{Peter A.\ Shaver}
\affil{ESO, Karl-Schwarzschild-Strasse 2, Garching, D-85748, Germany;
pshaver@eso.org}

\begin{abstract}
We use cosmological simulations of reionization to predict the effect of
large \HII\ regions around individual high-redshift quasars  on the possible
signal from the redshifted $21\dim{cm}$ line of neutral hydrogen in the
pre-reionization era. We show that these \HII\ regions appear as
``spectral dips'' in frequency space with equivalent widths in
excess of $3\dim{mK MHz}$ or depths in excess of about $1.5\dim{mK}$, 
and that they are by far the most prominent cosmological signals in
the redshifted \HI\ distribution.

 These spectral dips are expected to be present in almost every line 
of sight.
If the spectral dips of a large enough sample of \HII\ regions are
well resolved in frequency space, the
distribution of line depth and equivalent width in frequency with a
known observing beamsize can be used
to infer the \HII\ region size distribution and the mean difference in neutral hydrogen density between the \HII\ regions ( which may contain self-shielded neutral gas clumps) and the surrounding medium, providing a powerful test for models of reionization.
\end{abstract}

\keywords{cosmology: theory - cosmology: large-scale structure of universe
  - diffuse radiation - galaxies: formation - galaxies: intergalactic
  medium - radio lines: general}

\section{Introduction}

  Theoretical interest in the redshifted $21\dim{cm}$ line signal from
intergalactic hydrogen before and during the reionization era has
surged in recent years. Fluctuations in 
the $21\dim{cm}$ signal can be generated by variations in the density,
spin temperature, or ionized fraction of the intergalactic gas. These
could arise from primordial density fluctuations, from spatial
variations of the radiation intensity near the wavelength of the \lya\ 
hydrogen line (which affects the spin temperature), from other radiation
that can heat the gas, or from \HII\ regions created during the process
of reionization by discrete sources. The latter source of $21\dim{cm}$
fluctuations was discussed in the pioneering work of Scott \& Rees
(1990), Madau, Meiksin, \& Rees (1997), and Tozzi et al.\ (2000), but
has received only limited attention recently (Wyithe \& Loeb 2004).

However, ionized (\HII) regions around high-redshift quasars can easily
reach sizes comparable to the angular beams of future radio telescopes such as
LOFAR and SKA. Thus, they could be the most prominent features in the
angular and frequency distribution of the redshifted $21\dim{cm}$ signal
from the early universe, and the first features to be measured observationally.

In this paper we estimate the effects of these large features on
the possible cosmological signal by using numerical simulations of
reionization. This becomes feasible now as simulations of reionization
reach the level of sophistication and accuracy sufficient to model
quantitatively not only the small-scale details of reionization (Gnedin
2004), but also very large regions (up to a horizon size: Kohler, Gnedin, \&
Miralda-Escud\'e 2004). The latter approach is used here, because
quasars are relatively rare at high redshift, and sufficiently large
computational volumes (several hundred Mpc in size) must be used to include
a representative sample of quasars.

We concentrate solely on the redshifted $21\dim{cm}$ signal
in the frequency domain, because, as has been shown before, foreground
contamination is a severe limiting factor in observing the cosmological
features in the angular domain on the sky (Di Matteo et al.\ 2002; Oh \&
Mack 2004; Gnedin \& Shaver 2004). Several recent papers have discussed elegant
and sophisticated approaches to circumventing this contamination (Zaldarriaga, 
Furlanetto \& Hernquist 2004; He et al.\ 2004; Santos et al.\ 2004, 
Loeb \& Zaldarriaga 2004), but these approaches would require bigger 
telescopes than 
currently planned, so it is most likely that the first detections will
be made in the frequency domain and attempts to de-contaminate the 
angular signal will require follow-on observations.

\section{Simulation}

  In order to model the spectral signatures of \HII\ regions in the
early universe, we need to simulate a large enough region of space that
includes bright quasars during the reionization era. Such 
quasars are sparse at high redshift, since in the hierarchical
structure formation paradigm they only form in extreme over-densities
at early times. 

  The simulation code we use is a modified version of the
``Softened-Lagrangian Hydrodynamics'' code described in Gnedin (2004),
which follows the evolution of dark matter and gas. The code includes a
model for star formation in regions of dense gas that can cool, and for
the emission of ionizing radiation from stars. Radiative transfer is
fully included using the method described in Gnedin \& Abel (2001), and the
ionization state of the gas is followed in every cell as determined by the
local radiation intensity.

In this specific project we use
a multi-resolution approach that utilizes a small-scale simulation as a
``sub-cell'' model for larger-scale simulations. The ``sub-cell'' model is
implemented by using the so-called ``clumping factors'', which 
account for the structure on scales too small to be resolved in a
large-scale simulation. The complete description of our approach and the
appropriate tests are presented in a separate paper (Kohler et al.\ 2004).

 Using the clumping factor formalism allows us to increase the size of
the simulation box to several hundred comoving Mpc, while simultaneously
accounting for the ionizing radiation transfer and recombination rate in
a gaseous medium that is highly inhomogeneous on small scales.

\begin{figure}[t]
\plotone{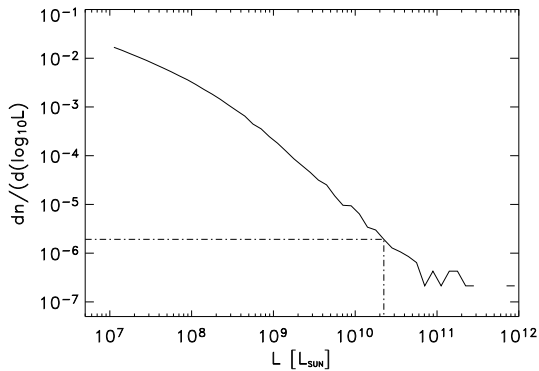}
\caption{Our adopted quasar luminosity function at $z=10$, as 
extrapolated from Schirber and 
Bullock (2003). The dash-dotted line shows the luminosity of a typical
quasar which is expected to be present in almost all $10^\prime$
observational beams.}
\label{fig:lumfun}
\end{figure}

 In our simulation of reionization, we add individual quasars as discrete, 
luminous sources of ionizing radiation in addition to the emission associated
 with star-formation. The luminosity distribution of quasars added to the 
simulation is chosen from the model of Schirber \& Bullock (2003). 
This model is consistent with all known
observational data on the evolution of the quasar luminosity function,
including recent GOODS data (Cristiani et al.\ 2004). The impact of quasars
 on reionization depends of course on the luminosity function at very high 
redshift, $z=6$, before reionization ended, so we have to extrapolate 
the Schirber \& Bullock model to redhifts higher than observed quasar redshifts, which introduces a substantial uncertainty. Our goal, therefore,
is to discuss a range of possibilities, as precise predictions are not
possible. 
The extrapolated luminosity function of quasars introduced in the simulation 
is shown in Figure~\ref{fig:lumfun} for 
$z=10$; the finite box size causes the discreteness effects observed 
at high luminosity end. 
Our simulation box 
 includes quasars up to a luminosity of about $10^{12}L_{Sun}$ at
 $z=10$
(although quasars this luminous are rare, and therefore appear in only
a few lines of sight).

In our simulation we include biasing of sources of ionizing and \lya\
radiation. Specifically, we assume that galaxies are biased with a
bias factor of two, while quasars are biased with a bias factor of
three. In this work we do not consider redshift-dependent bias factors.

  The specific simulation used in this paper has $128^3$ cells with an
average comoving cell size of $2h^{-1}\dim{Mpc}$. The box size of
$256h^{-1}\dim{Mpc}$ allows us to resolve individual \HII\ regions
around typical quasars at $z\sim8-10$, although it is still too small to
include the brightest high-redshift quasars found in the Sloan Digital
Sky Survey (Fan et al.\ 2003). But since we are interested in typical
(rather than the rarest, most extreme) features in the redshifted
$21\dim{cm}$ signal, this simulation is sufficient for our purpose.

\section{$21\dim{cm}$ ``Spectral Dips'' due to Quasar \HII\ Regions}

  We now investigate the characteristics of \HII\ regions as they appear
in the redshifted $21\dim{cm}$ spectra from neutral hydrogen. The
intergalactic hydrogen can be observed in absorption or emission against
the CMB, depending on whether the spin temperature is lower or higher
than the CMB temperature. Although an absorption signal is expected when
the earliest emission of \lya\ photons occurs (which couples the spin and
kinetic temperature of hydrogen, bringing the spin temperature below the
CMB temperature; see Madau et al.\ 1997), the 21 cm signal in our simulation
quickly turns into emission, as X-rays, far-ultraviolet radiation, and weak 
shocks from structure formation heat the atomic gas
to a temperature higher than the CMB (e.g., Chen \& Miralda-Escud\'e
2004; Gnedin \& Shaver 2004). For the redshift range that we consider
in this paper (corresponding to frequencies largely above the FM
bands), the atomic gas is 
heated above the CMB temperature and produces an emission signal.
Therefore, \HII\ regions should produce dips in the spectrum because
they contain less neutral hydrogen than their environment, so their
emission is reduced compared to the regions around them. We shall
generally refer to these \HII\ region features as dips. Note that these
dips look like absorption features even though they have nothing to do
with absorption.

  The basic observable properties of the spectral dip of an \HII\
region are its width and depth. The depth depends on the ionization
fraction, gas density, spin temperature, and also on the angular size of
the \HII\ region if this angular size is not resolved by the beam of the
radio telescope that is used (the angular size of a typical observable \HII\
region is about $1.5-2 \dim{arcmin}$). The width of the spectral dip in frequency
space is determined by the radial extent of the ionized region (assuming
that peculiar velocities are small compared to the Hubble expansion
across the \HII\ region).  

In practice, the detected cosmological \HII\ regions will probably
be smaller than the angular size of the radio beam. 
However, if the \HII\ regions are resolved in frequency, their frequency
width determines their size. Thus, the average linear (and angular)
size for a sufficiently large sample of observed \HII\ regions can be 
estimated (for a given cosmological model) from the frequency data
alone, and, taking into account the dilution by the known radio
beamsize, the actual average depth of the spectral dip can be inferred from
the measured dip depth. 

In this paper, however, we do not investigate such a measurement any
further, because it will most likely require a large number of
independent lines of sight. 

\hide{
  Let us assume that we have derived an average dip depth in this way
  (i.e. corrected for convolution by the radio beamsize), which we
  call $T_{B,obs}$. 
We can then compare this with the dip depth, $T_{B0}$, that would
be expected from \HII\ regions with a gas density equal to the mean
density of the universe, if the \HII\ region is completely ionized and
the surrounding medium is completely neutral. This expected dip depth
is given by
$$
T_{B0} \simeq 25\dim{mK}
\left( {\Omega_b h\over 0.03} \right)
\left( {0.3\over \Omega_m} \right)^{1/2}
\left( {1+z\over 10} \right)^{1/2}
$$
\begin{equation}
\phantom{AAA}\left( 1 - {T_{CMB} \over T_s} \right).
\end{equation}
Assuming again that the mean baryon density and matter density
($\Omega_b$ and $\Omega_m$) and Hubble constant
($H_0=100h\dim{km/s}/\dim{Mpc}$) are known, and furthermore that the spin
temperature $T_s \gg T_{CMB}$, the quantity
$T_{B0}$ is known at any redshift. The observed average dip depth $T_{B,obs}$ 
may differ from $T_{B0}$ because some fraction $x_{\HII}$ of gas in
\HII\ regions may be atomic. In particular, some of the dense gas
clumps associated with collapsed halos inside the \HII\ region can be
self-shielded and neutral, so the fraction $x_{\HII}$ depends on the
small-scale density structure of the intergalactic medium. At the same
time, the medium outside the \HII\ regions may have an atomic fraction
$x_{\HI}$ that is less than unity, if a set of smoothly distributed sources
has been ionizing the medium more homogeneously without detectable \HII\
regions. Then, the observed dip depth is
\begin{equation}
T_{B,obs} = T_{B0}\, (x_{\HI} - x_{\HII})
\label{tbobs}
\end{equation}
}

The average depth of a spectral dip is related to the difference in the
amount of neutral hydrogen present (on average) inside and outside an
\HII\ region. The observational
determination of this quantity would introduce a new tool for testing
models of reionization. In the \HII\ regions the fraction of gas that is 
still atomic depends on how much of the gas in collapsed, low-mass halos 
remains self-shielded, and how much has been ionized and evaporated from 
halos (e.g., Shapiro, Iliev, \& Raga 2004), a process that is important to 
determine the clumping factors of ionized gas. Outside the identified 
\HII\ regions, the fraction of gas already ionized depends on the 
presence of hard photons and low-luminosity sources producing unresolved 
ionized regions. Determining the dip depth distribution will require a 
large number of independent lines of sight, and the theoretical
 interpretation will be complex. In the rest of this paper we focus
 on how spectral dips due to \HII\ regions can be identified.

\hide{
  We note that an additional complication arises because the first
sources of ionizing radiation should appear in high-density regions, so
\HII\ regions should also be found in regions of high density, when
the density is smoothed over a scale comparable to the \HII\ region
radius. Therefore, an additional uncertainty in equation (\ref{tbobs})
is the factor by which the mean density of \HII\ regions may be higher
than the mean gas density of the universe.
}

  In the absence of \HII\ regions, the distribution of brightness
temperature of the redshifted $21\dim{cm}$ signal should be Gaussian,
reflecting the 
imprints of cosmic density fluctuations, which are linear on the large
scales that are most easily observable.
The \HII\ regions add additional features in
the spectrum that are mainly caused by variations in neutral fraction.
Gaussian regions of lower
emission correspond to regions of lower density, but \HII\ regions usually have
higher than average density and a dramatically lowered neutral
fraction. This can be used for distinguishing Gaussian fluctuations from
those caused by the ionization from quasars in the simulation, but in
real observations we will have to rely on the intrinsic properties of
spectral dips to separate \HII\ regions from density fluctuations. In
the next two sections we will use our simulation to predict how such a
separation can be made from the observational data.

\section{Analysis of Simulated Lines of Sight}

  We now discuss how we use the output at consecutive timesteps from the
simulation described in \S 2 to create synthetic spectra of redshifted
$21\dim{cm}$ emission. These spectra reflect the large-scale
distribution of gas density, spin temperature, and hydrogen neutral
fraction. The spin temperature is computed self-consistently in the
simulation including the effects of \lya\ pumping and collisions
with electrons and neutral atoms
(see Gnedin \& Shaver 2004 for details).
The pixel size in the spectra is set by the cell size of
$2h^{-1}\dim{Mpc}$, which 
corresponds to a frequency range of about $0.2\dim{MHz}$ at $z\sim9$.

  The synthetic spectra are created by casting a random line of sight
through the simulation box. Our simulated lines of sight span a
significant fraction of the horizon distance,
so we simulate the line of
sight on a fixed past light-cone instead of a fixed cosmic time,
using consecutive output files at different timesteps of the simulation.
We also account for a finite observing beamsize by convolving 
the 21cm signal over a plane perpendicular to the line of sight with a
Gaussian beam, with a physical radius that varies over the line of
sight according to a fixed angular size of a radio beam $\Theta_b$.

  The starting redshift of our spectra is $z=13$, which corresponds to a
frequency of $101\dim{MHz}$. This redshift was chosen to
place the resulting frequency in the range of next-generation radio
observatories.
We used 450
independent lines of sight to accumulate sufficient statistics.

  After the synthetic spectra were created in this way, we analyzed
the brightness temperature distribution for signatures of \HII\ regions.
The process we followed for this analysis consists of the following steps.
First, we fit the mean signal $\langle T_{B}\rangle$
with a fourth-degree polynomial in order
to remove variations on a very wide frequency scale. This process,
similar to the continuum fitting used in the analysis of \lya\ forest
spectra, is intended as an approximation to the removal of the
combined galactic and extragalactic foregrounds and the cosmological
mean signal in an observation. Next,
we calculate the fluctuating signal as:
\begin{equation}
  \Delta T_{B} = T_{B} - \langle T_{B}\rangle,
\end{equation}
to obtain a fluctuating signal with zero mean that can be searched for
\HII\ region dips. For the analysis we disregard all parts of the spectrum in which the
emission exceeds the mean signal, which correspond to regions of high
gas density that are part of linear Gaussian fluctuations. 
We are then left with a spectrum showing dips
due to both Gaussian fluctuations (corresponding to regions of
low gas density) and \HII\ regions around bright
sources (corresponding to regions of very low neutral fraction compared
to the average medium).

\begin{figure}[t]
\epsscale{0.90}
\plotone{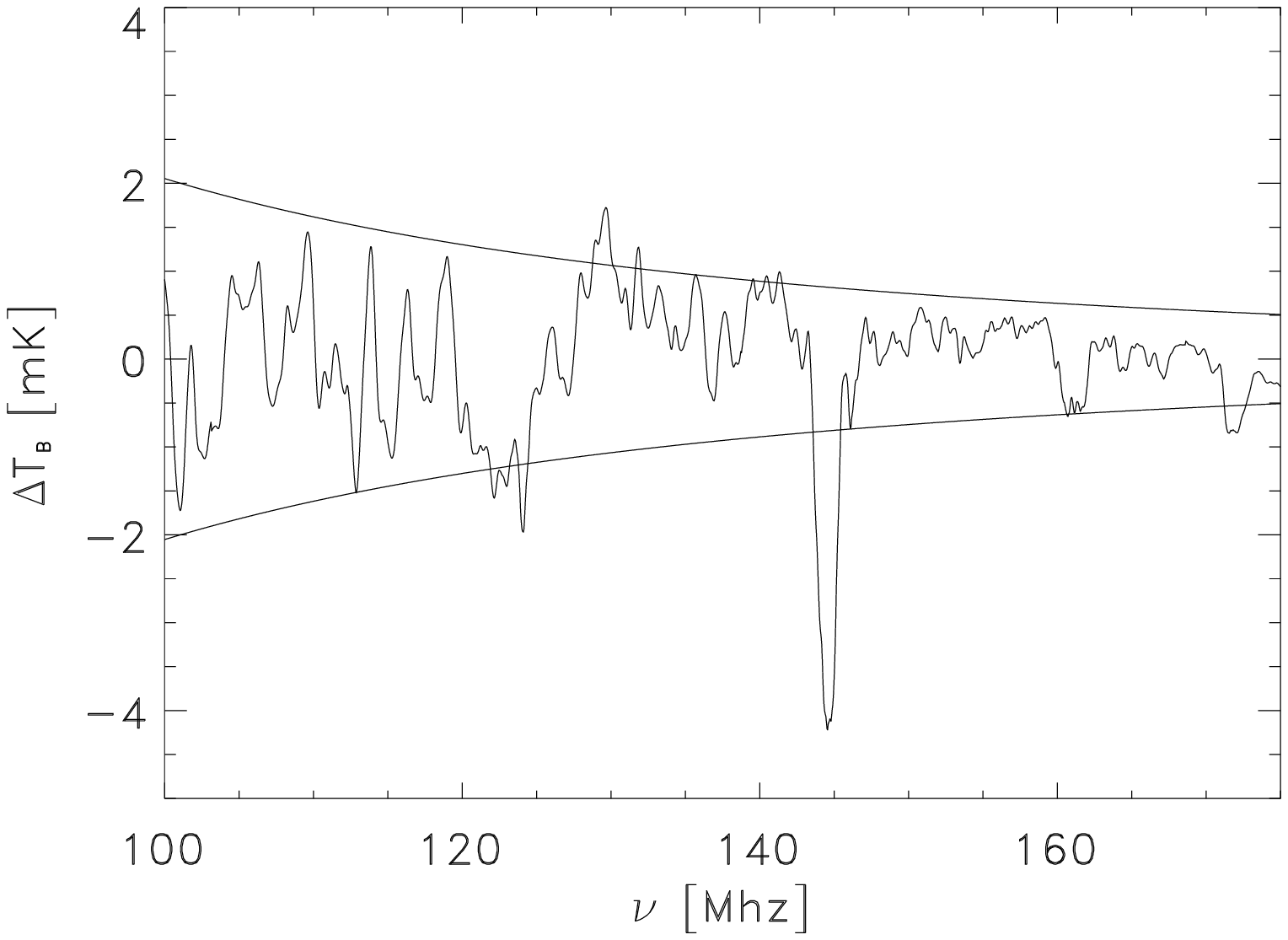}
\plotone{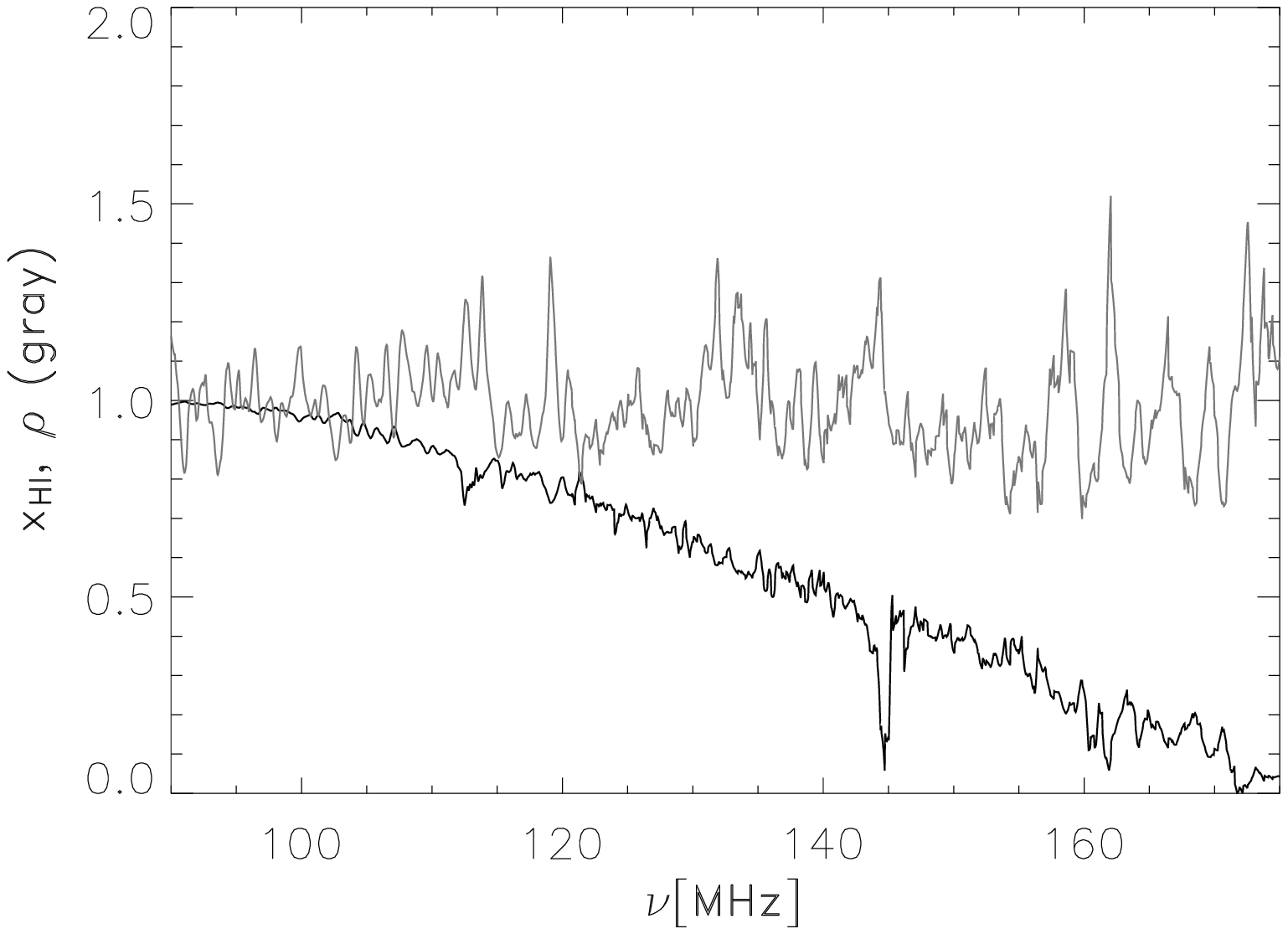}
\caption{Top panel: the fluctuation in the
brightness temperature $\Delta T_{B}$ as a function 
of frequency in a $10^\prime$ beam with a $0.5 \dim{MHz}$ bandwidth. 
Also shown are the $5 \sigma$
sensitivity limits for a $1000\dim{hrs}$ integration time (Shaver
et al.\ 1999).  Bottom panel: the distribution of gas density (gray)
and hydrogen neutral fraction (black) along the same line of sight. 
The \HII\ region spectral feature can be seen at $\nu=144\dim{MHz}$.
}
\label{fig:spectrum}
\end{figure}

\section{Results}

In Figure~\ref{fig:spectrum} we show an example of a typical line 
of sight with
the mean signal subtracted. The top panel shows the 
brightness temperature fluctuation $\Delta T_{B}$ versus $\nu$ after
subtracting  
the mean signal. A significant fluctuation due to an \HII\ region at 
$\nu=144\dim{MHz}$ is easily identifiable in the spectrum. 
Superimposed are the $5 \sigma$ sensitivity limits for an integration time 
of $1000\dim{hrs}$ and a bandwidth of $0.5 \dim{MHz}$ (we discuss our
choice for the bandwidth below).

It is important to note here that our results differ to a small degree from
the predictions of Tozzi et al.\ (2000) and Wyithe \& Loeb (2004) in that 
we do not observe an
increase in the redshifted $21\dim{cm}$ emission just outside the \HII\
region. This is due to the 
fact that the main source of \lya\ radiation that couples the gas
kinetic temperature and the spin temperature of the $21\dim{cm}$ transition
in our simulation is normal galaxies. Since,in our simulation,  galaxies
 typically form before
the quasars, an \HII\ region
around a typical high-redshift quasar expands in the already pre-heated
gas, in which the spin temperature of the $21\dim{cm}$ transition is
already much higher than the CMB temperature, so the extra \lya\
radiation from the quasar does not lead to an increase in the emission. We
would like to re-emphasize here that, since we are treating the radiative
transfer in full 3D, we do include the proximity effect of the increased
\lya\ radiation around quasars, but this effect, as our results
indicate, is not significant.

To distinguish \HII\ regions from under-dense regions with low 
neutral hydrogen density, we now look at the distribution of
density and neutral fraction in the same frequency range in the data 
obtained from the simulation. 

The bottom panel
in Figure~\ref{fig:spectrum} depicts the variations in these two
parameters. The neutral fraction shows
a dip at $\nu=144\dim{MHz}$, whereas the density shows a definite
increase. Consequently we can determine that this fluctuation results from
a high density region containing more ionized gas than the average. We
expect to find quasars in high density regions since their population is
biased. If the radiation intensity were approximately uniform, we would expect
 to find lower densities of neutral gas
in under-dense regions of the IGM, because of the decreased recombination
rate. Here the high density and ionized fraction of the IGM allow us to
conclude that the spectral dip is caused by an \HII\ region surrounding
a bright quasar. 

\begin{figure}[t]
\plotone{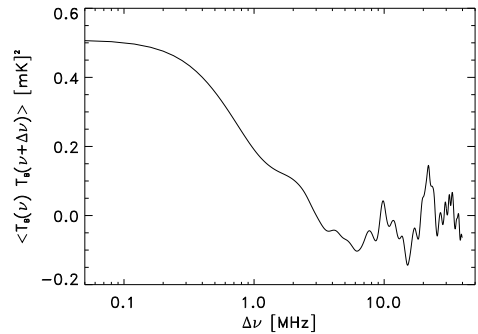}
\caption{The auto-correlation function of the brightness 
temperature $\Delta T_{B}$ as a function of FWHM frequency width
$\delta\nu$.}
\label{fig:correlation}
\end{figure}
The fluctuations in the brightness temperature on large scales are
expected to be Gaussian, reflecting the large-scale variations in the
cosmic gas density and neutral fraction. However, on sufficiently
small scales the fluctuations become correlated. Thus, unless the beam
size of a radio telescope is small enough (less than about 1 arcmin)
to resolve individual \HII\ regions around high redshift galaxies, one
would expect the cosmological signal to become smooth on scales
comparable to the correlation length of high redshift galaxies. Figure
\ref{fig:correlation} shows the auto-correlation length of the
brightness temperature from the line of sight shown in Fig.\
\ref{fig:spectrum} 
Thus, decreasing the bandwidth of observation to below
about $0.5\dim{MHz}$ will not result in measuring new information in
the signal unless it also coincides with a decrease in
the beam size (this is also apparent from Fig.\ 11 of Gnedin \& Shaver 2004).
In the rest of this paper, we assume a fixed value of $0.5\dim{MHz}$
for the bandwidth for our fiducial beam size of $10^\prime$.

\begin{figure}[t]
\plotone{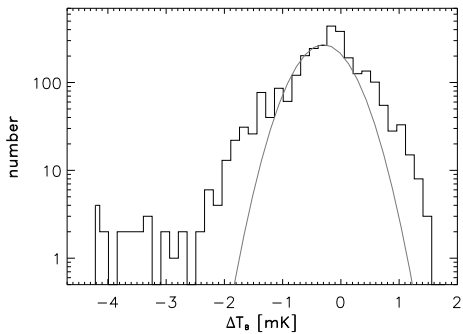}
\caption{The histogram of $\Delta T_{B}$ for the normalized spectrum from Figure~\ref{fig:spectrum}: notice
  the 
  long  tail of the non-Gaussian contribution from \HII\ regions. }
\label{fig:histogramTB}
\end{figure}
In addition to investigating individual spectra as shown above, we can
use the variations of
brightness temperature around the average to look at the distribution of
spectral dips statistically. 
Figure~\ref{fig:histogramTB} shows a histogram of the
difference in brightness temperature of the spectrum in
Figure~\ref{fig:spectrum}. One can see the tail of the distribution on the
negative $\Delta T_{B}$ side that correspond to dips in the spectrum due
to quasar \HII\ regions. 
The peak centered around $\Delta T_{B} = 0 $ is
mainly due to the Gaussian fluctuations, but the long tail on the
negative $\Delta T_{B}$ reaching  $\Delta T_{B} = -4.5 \dim{mK}$ shows
the significant contribution from regions that are due to ionized
regions and not from Gaussian density fluctuations.

\begin{figure}[t]
\plotone{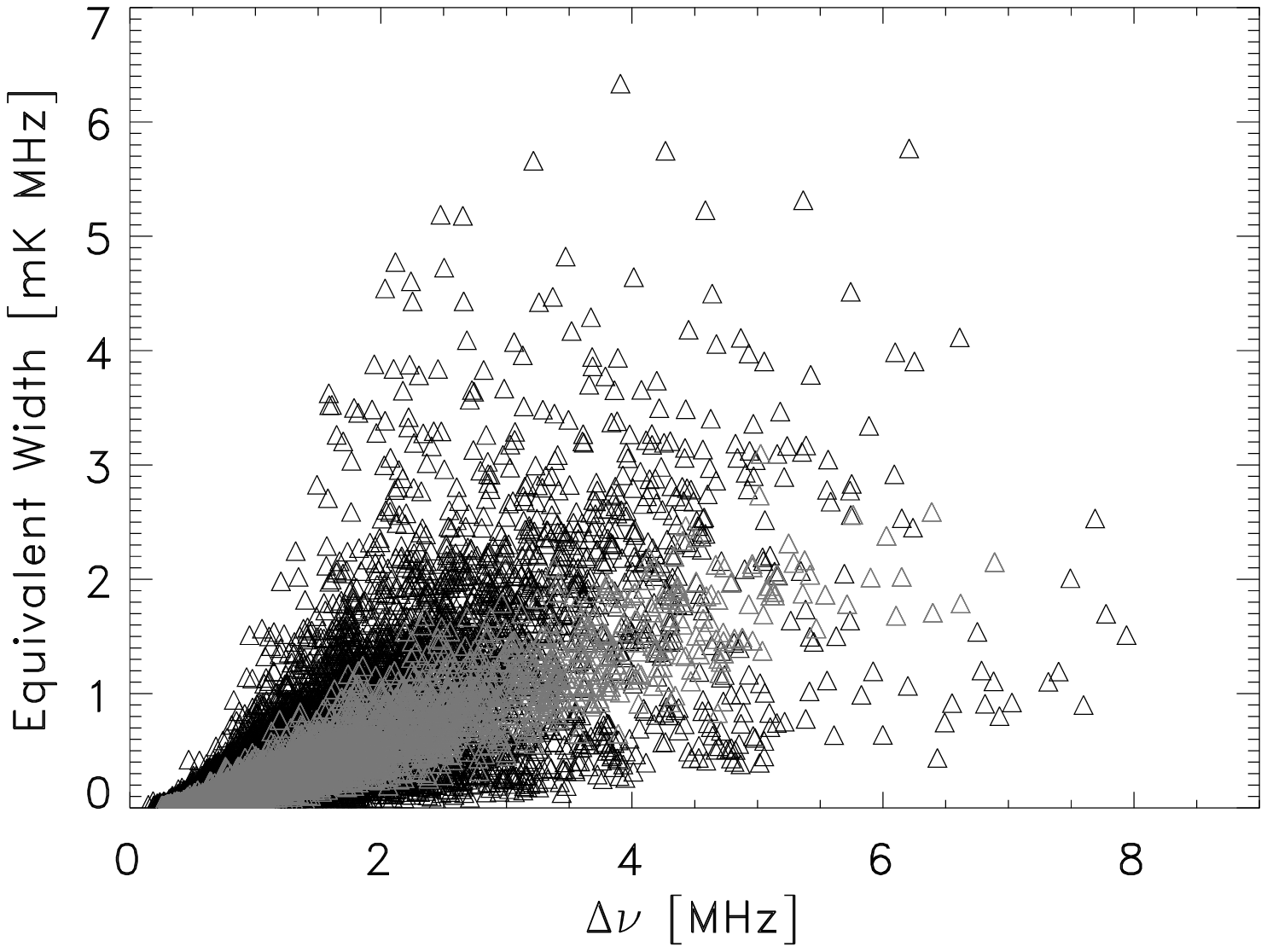}
\plotone{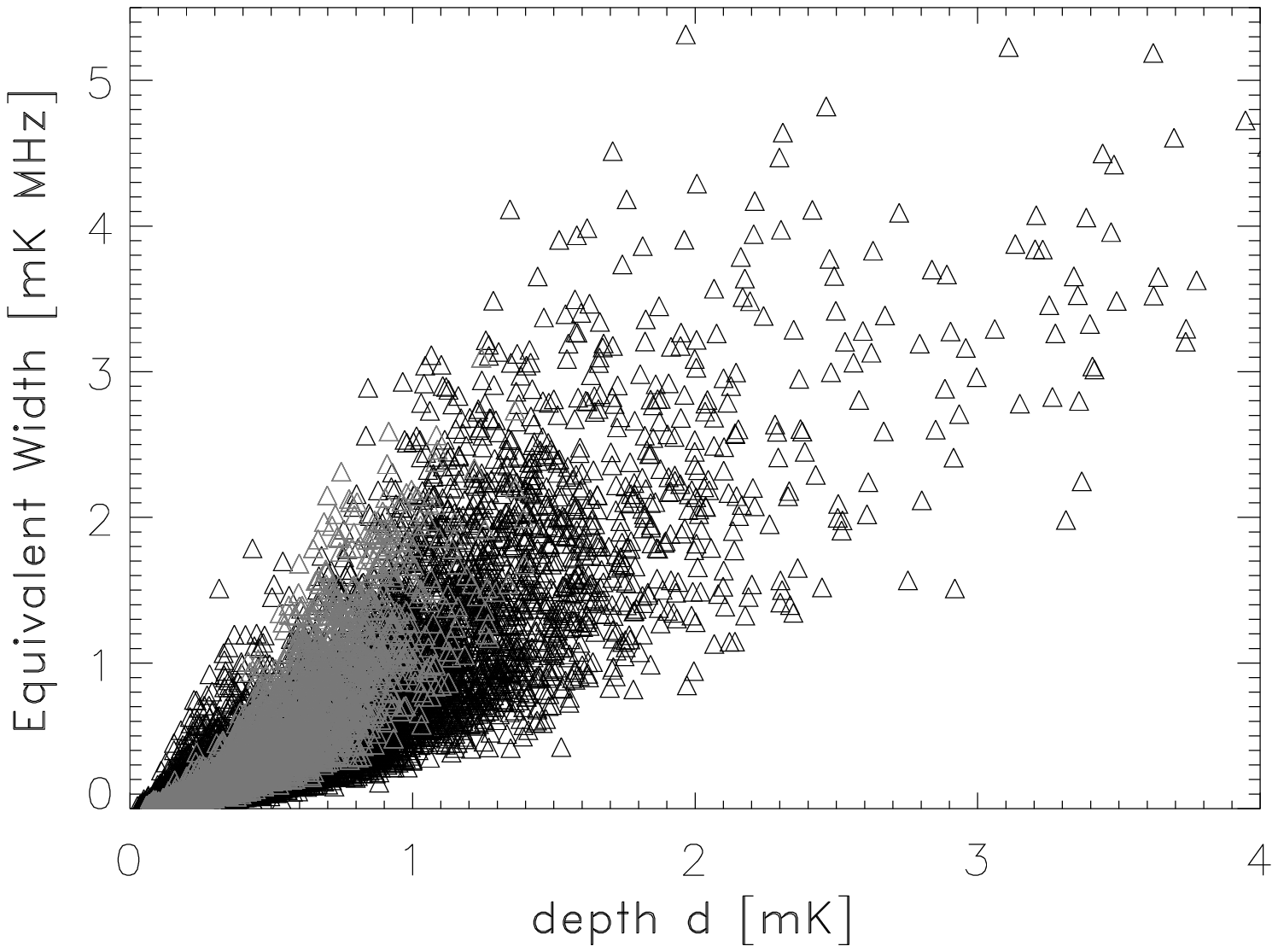}
\caption{Top panel: in $\textit{black}$ is shown the distribution of
  equivalent widths and sizes of the spectral dips from 450 lines of
  sight. The
  $\textit{gray}$ data shows the same distribution for a pure 
  Gaussian distribution. Bottom panel: A similar plot but for
  equivalent width versus the depth of the spectral dip distribution.}
\label{fig:gaussianequwidth}
\end{figure}

  We use a simple calculation of the equivalent width of each dip as a
way to measure the dip depth. We integrate the area below the mean
signal for each dip (i.e.\ a connected region with $\Delta T_B<0$), 
and compile a distribution for all the dips of 450
different spectra. This distribution is shown in
Figure~\ref{fig:gaussianequwidth} by plotting the equivalent width versus
the dip width in frequency space (measured as FWHM). 
As a comparison, we also show
the same distribution for artificial
lines of sight that contain no \HII\ regions, so
that any dips are entirely due to Gaussian variations in
cosmic density.

It is clear that the largest equivalent widths and depths 
correspond to \HII\ regions
around quasars, since the Gaussian density 
variations do not generate such large
dips. The radial sizes, as shown by $\Delta \nu$, of both samples are
comparable, since the Gaussian density
fluctuations also yield the wide dips seen
in the spectra. However, the Gaussian fluctuations are much shallower,
as can be seen by 
the smaller values in the equivalent width. 
There is also a range of smaller equivalent
widths that is not reached with purely Gaussian fluctuations. This could be
due to faint quasars in the simulation causing the small features. 
Judging from this graph, all spectral dips with
equivalent width in excess of $3\dim{mK}\dim{MHz}$ or depth in excess
of $1.5 \dim{mK}$ are
caused by individual \HII\ regions. 

\begin{figure}
\plotone{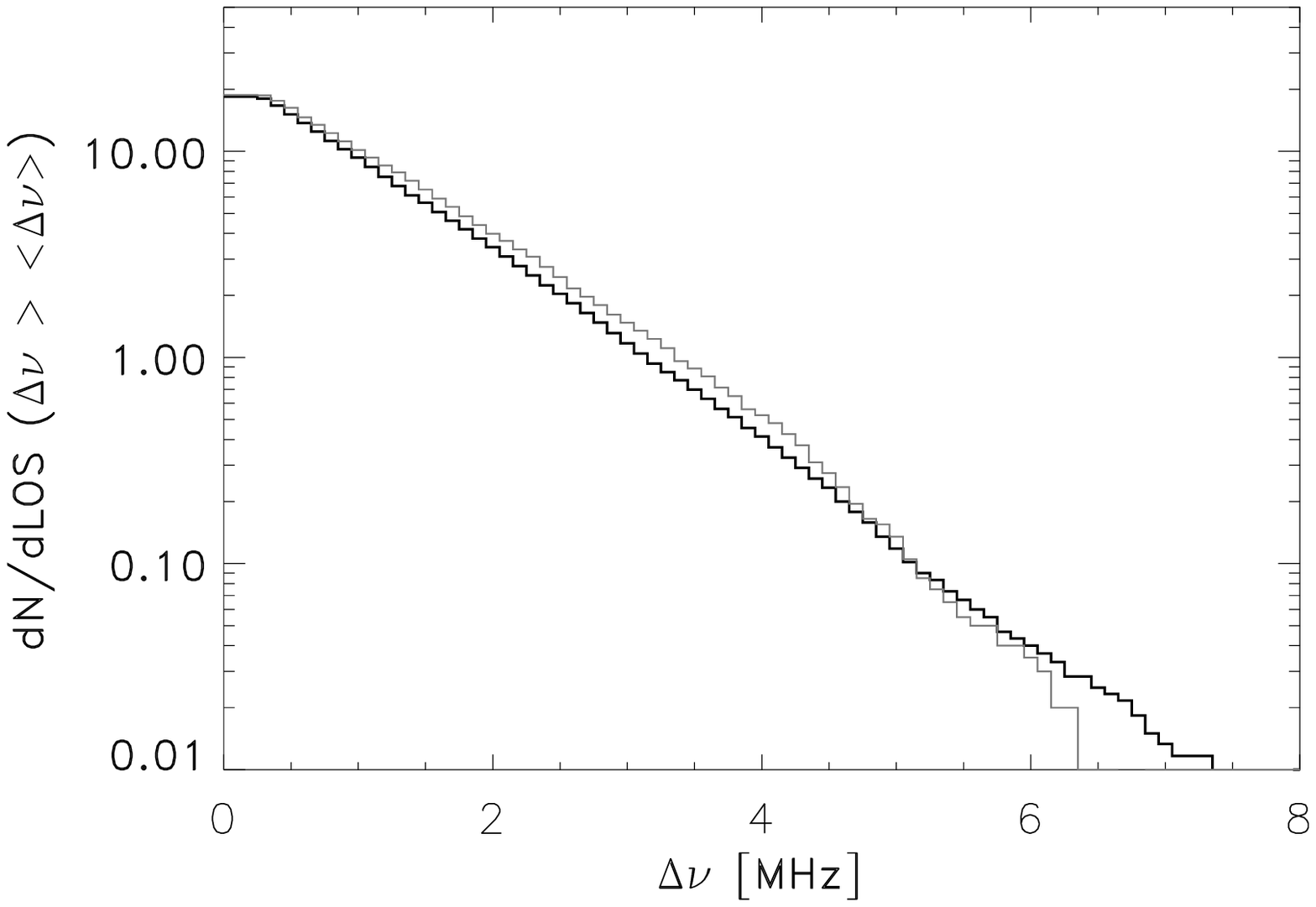}
\plotone{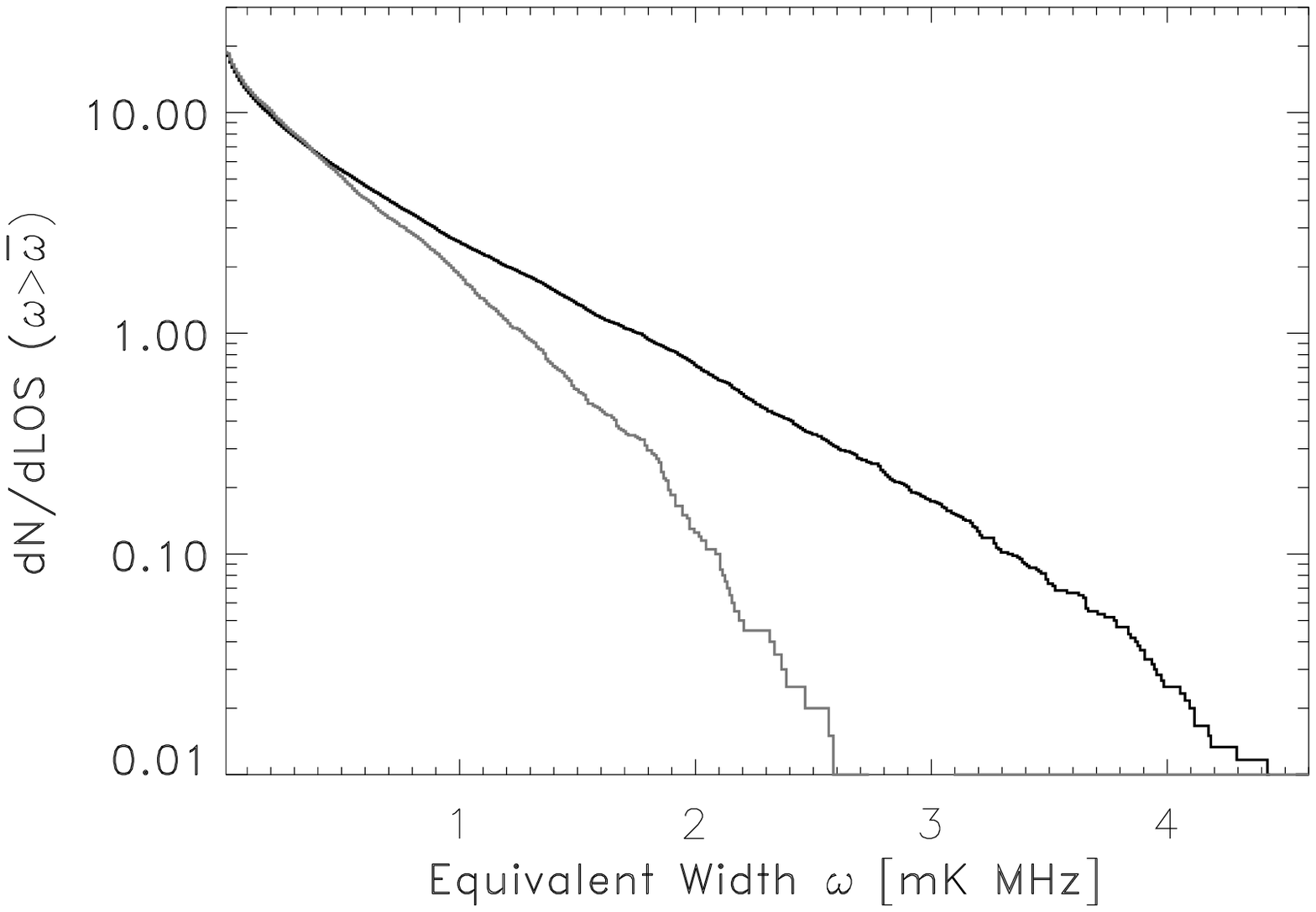}
\end{figure}

\begin{figure}
\plotone{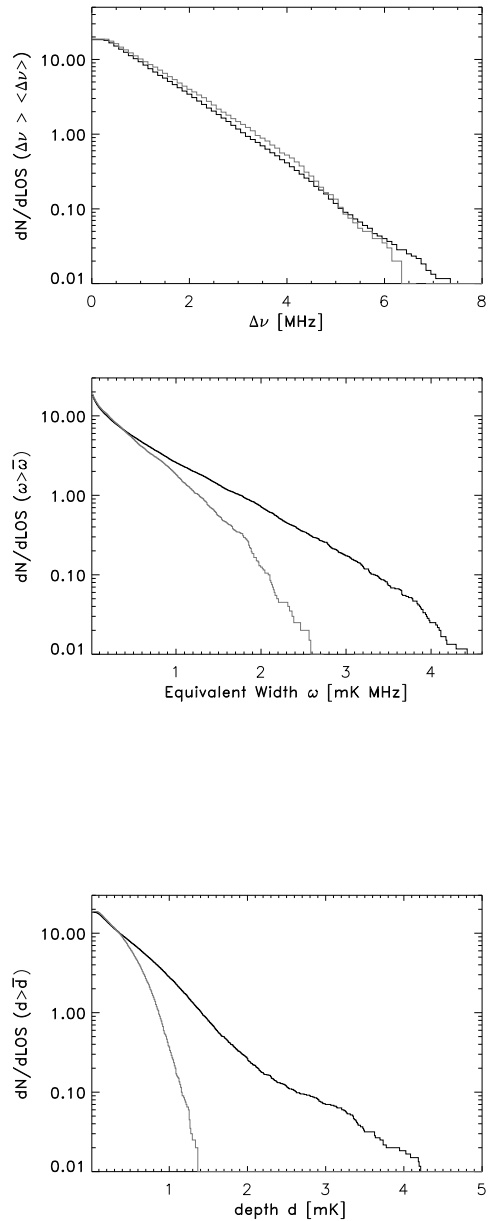}
\caption{
  Top: The cumulative distribution of spectral dips per line
  of sight as the function of the dip width $\Delta\nu$ (measured
  as FWFM - {\it top\/}), the equivalent width ({\it middle\/}), and
  the depth ({\it bottom}) for the simulated spectra (\textit{black})
  and for purely Gaussian fluctuations (\textit{gray}).}
\label{fig:histtheta}
\end{figure}

We can also look at cumulative distributions of various parameters
characterizing these spectral dips, which are shown in
Figures~\ref{fig:histtheta},  
to determine which are due to Gaussian fluctuations
compared to the ones due to \HII\ regions. The figure shows distributions
for the simulated fluctuations in black and contrasts it with the
distribution for pure Gaussian fluctuations in gray. 
From Figure~\ref{fig:histtheta}a it is again clear that all spectral dips
with equivalent widths in excess of about $3\dim{mK}\dim{MHz}$
or depths in excess of about $1.5\dim{mK}$ are coming from high-redshift
\HII\ regions. The widths of the dips in frequency space, however,
cannot be used to separate \HII\ regions from linear Gaussian fluctuations.
From these graphs we can infer that there is about a one
in three chance to detect an \HII\ region with equivalent width in
excess of $3\dim{mK}\dim{MHz}$ in each line of sight, but narrower deep
dips (with depth in excess of $1.5 \dim{mK}$) are present in 
almost any line of sight. 

The results shown in Figures~\ref{fig:histtheta} were obtained for a 
bandwidth of $0.5 \dim{MHz}$, but as long as the bandwidth is less than
about $1\dim{MHz}$, spectral dips from \HII\ regions are resolved (as
can be seen from Fig.\ \ref{fig:gaussianequwidth}a), 
and the distributions of equivalent widths and 
depths remain the same.

\section{Discussion}

We have shown that by far the easiest to observe features of the cosmological
signal of redshifted $21\dim{cm}$ emission from neutral hydrogen in the
pre-reionization era are the spectral dips due to individual
\HII\ regions from high-redshifts quasars. The largest of these features
can be easily distinguished from Gaussian fluctuations due to
large-scale structure: for
example, we find that all
spectral dips with equivalent width in excess of
$3\dim{mK}\dim{MHz}$ or depth in excess of $1.5\dim{mK}$ come exclusively from
quasar \HII\ regions (for a $0.5\dim{MHz}$ bandwidth).

 These dips are not rare - for our adopted plausible (albeit
uncertain) extrapolation of the quasar luminosity function, detectable
dips are present in almost every line of sight. This
conclusion may sound counterintuitive to the common wisdom that ``high
redshift quasars are rare'', but we need to emphasize that the 
quasars that create detectable \HII\ regions are much less luminous (and 
much more numerous) than, say, SLOAN quasars. Besides, a $10^\prime$
beam (for example) going from $z=6$ to $z=12$ (the onset of the FM
radio band) contains a significant volume of about 
$500{,}000(h^{-1}\dim{Mpc})^3$. According to our model of the quasar
luminosity function, there is about one quasar within this volume with
blue rest-frame luminosity of $2\times10^{10} L_{Sun}$, and this
luminosity is sufficient for a typical quasar to create an \HII\
region that can be distinguished from Gaussian fluctuations due to
large-scale structure. For a different model of the quasar luminosity
function, our results can be simply scaled with the abundance of
quasars above this characteristic luminosity.

Knowing the shapes and sizes of the ionized regions around quasars at high
redshift could help to determine better the physical processes near
super-massive black holes at the very first stages of structure formation
in the universe. In particular, if observed spectral dips in the
redshifted $21\dim{cm}$ emission can be identified with specific
quasars observed in the infrared by, say, JWST, then a significant
constraint can be placed on the lifetimes
of quasars as a function of their luminosity. Such a constraint
is difficult to obtain using other methods. 

The depth distribution of spectral dips caused by \HII\ regions can put a 
constraint on the residual neutral fraction inside \HII\ regions. 
This neutral fraction would likely be locked in high-density, 
self-shielded gas clumps in collapsed halos (observed as Lyman 
limit systems in ultraviolet spectra), which play the dominant role 
for determining the clumping factor of ionized gas during reionization.

Under the assumption that the distribution of \HII\ regions is
statistically isotropic, it will be possible to put
constraints on the residual neutral fraction inside \HII\ region, and,
since this neutral gas most likely is locked inside Lyman-limit
systems, an important constraint can be put on the abundance of low
mass objects.

In conclusion, the spectral features predicted in this paper - the
spectral dips due to \HII\ regions around individual high-redshifts
quasars - will be the most prominent signals in the redshifted
$21\dim{cm}$ line of neutral hydrogen from the pre-reionization
era. Their spectral sharpness will easily distinguish them from the
smoothly varying continuum spectra of the foreground emissions, and
their strength should make them detectable in reasonable integration
times with the new generation of low frequency radio telescopes.

\end{document}